\documentclass[12pt]{article}
\usepackage{latexsym}
\usepackage{epsfig}

\usepackage{graphicx}
\usepackage{epstopdf}

\usepackage{epsfig,amssymb,amsmath,amscd}

\newcommand{\bq}{\begin{equation}}
\newcommand{\eq}{\end{equation}}
\newcommand{\bqa}{\begin{eqnarray}}
\newcommand{\eqa}{\end{eqnarray}}
\newcommand{\ben}{\begin{enumerate}}
\newcommand{\een}{\end{enumerate}}
\newcommand{\bc}{\begin{center}}
\newcommand{\ec}{\end{center}}
\newcommand{\bqb}{\begin{eqnarray*}}
\newcommand{\eqb}{\end{eqnarray*}}

\def\pr#1#2#3{Phys. Rev. ${\bf{#1}}$, #2 (#3)}

\def\pl#1#2#3{Phys. Lett. ${\bf{#1}}$, #2 (#3)}

\def\np#1#2#3{Nucl. Phys. ${\bf{#1}}$, #2 (#3)}

\def\jmp#1#2#3{J. Mod. Phys. ${\bf{#1}}$, #2 (#3)}

\begin{document}
\pagenumbering{arabic}
\thispagestyle{empty}
\def\thefootnote{\fnsymbol{footnote}}
\setcounter{footnote}{1}

\vspace*{2cm}
\begin{flushright}
July 23, 2018\\
 \end{flushright}
\vspace*{1cm}

\begin{center}
{\Large {\bf Z polarization in $e^+e^-\to ZWW $ 
for testing special interactions of massive particles.}}\\
 \vspace{1cm}
{\large F.M. Renard}\\
\vspace{0.2cm}
Laboratoire Univers et Particules de Montpellier,
UMR 5299\\
Universit\'{e} de Montpellier, Place Eug\`{e}ne Bataillon CC072\\
 F-34095 Montpellier Cedex 5, France.\\
\end{center}

\vspace*{1.cm}
\begin{center}
{\bf Abstract}
\end{center}

We show that the $ZWW$ production process may give complementary
informations about scale dependent heavy particle masses and
possible final state interactions as compared to previously studied
top quark production processes. We illustrate the $p_Z$ distribution
of the rate of longitudinal $Z_L$ component showing its sensitivity
to these effects which may arise from heavy particle substructure
or a dark matter (DM) environment.

\vspace{0.5cm}

\def\thefootnote{\arabic{footnote}}
\setcounter{footnote}{0}
\clearpage

\section{INTRODUCTION}

Our basic motivation is the search for simple signals of the existence of special
properties of heavy particles (t,Z,W,H,...) like a scale dependent
mass \cite{trcomp,CSMrev} or of final state interactions in multiparticle production  
(for example like in the hadronic case due to a substructure 
\cite{comp, Hcomp2,Hcomp3,Hcomp4,partialcomp}
or to a dark matter environment \cite{revDM,DMmass, DMexch}).\\
Previous papers \cite{eettZ,ggttZ,Wtb} were devoted to $Zt\bar t$ and $Wt\bar b$
production where the "simple signal" is given by the longitudinal gauge boson ($Z_L$ or $W^{\pm}_L$) rate of production. In SM it is, at high energy, equivalent to the rate of Goldstone boson $G^{0,\pm}$ production (up to corrections of order $m^2_{Z,W}/s$) \cite{equiv} for which the couplings to the particles involved in the considered process are indeed proportional to their masses. 
So finally the rate of longitudinal gauge
boson would give a measurement of the effective mass of the acompanying particles.
We have illustrated these properties for a set of interesting processes
$e^+e^-,\gamma\gamma,gg\to Zt\bar t, W^-t\bar b$ showing the sensitivity
to the top mass and also to a final state ($Z_Lt$ or $W_Lt$) interaction.\\ 
 However in these processes the sensitivity to the top mass arises
(as we can directly see from the Goldstone couplings) from the ratio
$m_t/v$ or $m_t/m_W$. So one should consider the possibility of
a simultaneous modification of the top and of the $W$ (and $Z$)
masses by scale dependence such that the resulting effective coupling
is only weakly or not at all affected.\\
It is therefore important to look at effects for different mass combinations. 
This is the motivation for the present paper and the analysis
of the $Z$ polarization in the $e^+e^-\to ZWW$ process where only
the $W$ and $Z$ masses are involved.\\
In Section 2 we present the basic SM properties and the details of
the sensitivity to these masses. In Section 3 we consequently show the effects
of a scale dependence of the heavy masses and we compare them to the case of the top quark in the previous processes. In Section 4 we also look at the effects
of final state interactions among $Z_L$ and $W_L$ states which may have some similarities with those of the masses.
Finally we summarize the possible implications of such observations
and how other processes could help about our aim.\\

\section{SM properties}

We will first look at the sensitivity of the $e^+e^-\to ZWW$ process to the values
of the $Z$ and $W$ masses.\\
In SM the Born diagrams are depicted in Fig.1. Each of these diagrams has a specific
dependence due to its kinematical feature and to the involved couplings. In addition
longitudinal polarization vectors have the peculiar $1/m$ dependence
\bq
\epsilon_L=({p\over m},{E\over m} \hat p)
\eq
An important SM feature is the equivalence of $Z_L$ amplitudes with the
Goldstone boson $G^{0}$ ones in $e^+e^-\to G^{0}WW$. The corresponding diagrams
are drawn in Fig.2. Apart from simple kinematical dependences, the mass
dependence appear in the  $H-WW$ and $\gamma,Z-GW$ couplings:

\bq
g_{HWW}={em_W\over s_W}~~~~~~~g_{\gamma GW}=em_W~~~~~~~g_{ZGW}=-e{m_Ws_W\over c_W}
\eq

 In a first study we will assume that the $m_W/m_Z$ ratio (i.e. $c_W$) is fixed.
This leads to  $G^{0}WW$ amplitudes proportional to the W mass. But the
$G^{0}W_LW_L$ amplitudes get, in addition, $1\over m^2_W$ factors which make them dominant at high energies and finally behaving like $1\over m_W$.\\

Apart from various $m^2_{Z,W}/s$ corrections (arising for example from
the difference between $p^0$ and $p$) these $Z_LWW$ amplitudes should be equivalent to the $G^0WW$ ones.\\

In practice this equivalence results from
cancellations at high $p_Z$ of several terms which individually
explode like $p/m$. Except for the $H$ exchange which is directly
equivalent to the corresponding one in the $G^0$ case and for the 2 diagrams of
the second set in Fig.1 which almost cancel each other, the cancellation
requires the addition of all other diagrams.\\

We will now compute the longitudinal $Z_L$ rate of production

\bq
R_L={\sigma(Z_L WW)\over \sigma(Z_T WW)+\sigma(Z_L WW)}
\eq
and the similar  $R^G_L$ one with $Z_L$ replaced by $G^0$.

They are illustrated in Fig.3 for $\sqrt{s}=5$ TeV and $\theta_Z={\pi\over3}$
and ${\pi\over2}$, versus $p_Z$, where one can see the importance of the
$m^2_Z$ terms in the accuracy of the equivalence.\\

\section{Sensitivity to scale dependent $W,Z$ masses}

We now want to see how this $m_W$ dependence will affect the
observables when one replaces the fixed mass value by some
scale dependence which may originate as mentioned in the
introduction by some substructure (like in the hadronic
case) or by some DM environment.

As in the previous analyzes we will use an effective scale dependence form

\bq
m_W(s)=m_W{(m^2_{th}+m^2_0)\over (s+m^2_0)}
\eq
\noindent
with $m_0=2,4$ TeV, leading to curves $"m2"$ and $"m4"$ in the
illustrations of the effects on the rates $R_L$ and $R^G_L$.

In the $G^{0}WW$ case the above mentioned $1\over m_W$ behaviour
of the leading $G^{0}W_LW_L$ amplitudes is responsible for the results shown
in Fig.4a,b.\\

First note the drastic difference with the top quark mass cases in 
($Ztt,Wtb$) production, \cite{eettZ,ggttZ,Wtb},
where the effects were exactly opposite.\\

The more subtle dependence of the $Z_LWW$ case is illustrated
in Fig.5a,b. As explained above the result is a mixture of the effects of
couplings, cancellations of exploding terms and $m^2_{Z,W}/s$ 
kinematical corrections.
The main features of the Goldstone case are nevertheless reproduced with
larger differences at low $p_Z$.\\

Note however that the Goldstone equivalence may not be
valid beyond SM except for special models designed to preserve
it.\\

\section{Final state interactions}

Independently of the possible presence of a scale dependent mass
there could exist strong final interactions essentially between
longitudinal $Z_L,W_L$ states. This may be due to their substructure
(like in the case of hadrons) or to a DM environment.\\
We have already considered this possibility in $Zt\bar t$ and $Wt\bar b$
production, \cite{eettZ,ggttZ,Wtb}. In the $ZWW$ case the 2 effects,
effective mass and final state interaction, respectively shown  
in Fig.5 and in Fig.6, may be cumulative.\\
In our illustrations we will use the same final state forms as
in the previous cases simply modifying the $Z_LW^+_LW^-_L$
amplitudes by the $(1+C(s_{ZW^+})) (1+C(s_{ZW^-})) (1+C(s_{W^+W^-}))$ 
"test factor"
with
\bq
C(x)=1+{m^2_{Z}\over m^2_0}~ln{-x\over (m_Z+m_W)^2} ~~, \label{Fs}
\eq
\noindent
with the subenergies $x=s_{ZW^+}$, $s_{ZW^-}$ or $s_{W^+W^-}$and $m_0=0.5$ TeV, like in \cite{DMexch}.\\
The resulting $p_Z$ distributions are shown in Fig.6a,b.\\

In order to distinguish these effects from the ones of effective
scale dependent masses detailed studies of their kinematical structures
may have to be done; for example the subenergies and angle
dependences. For this aim precise dynamical models should be considered.

\section{Conclusions and prospectives}

In this paper we have pursued our studies of possible effects
of scale dependent masses (for example due to
substructures) and of special interactions among heavy particles also 
generated by substructures or by a DM environment.\\

Previously we had essentially considered the case of the
top quark, the effect of $m_t(s)$, its influence on the accompanying
$Z_L$ or $W_L$ rate assuming no modification of the
Higgs mechanism producing the Z,W masses and generating these $Z_L, W_L$ states.\\

However if these $Z_L, W_L$ states are also affected, then the ratios $m_t/m_W$
controlling the $Z_L, W_L$ (or Goldstone) couplings to the top
quark may be very differently (weakly or even not at all) modified.\\

In such a situation the study of the $ZWW$ production processes could 
give essential and even decisive complementary informations as they 
will only depend on the $Z,W$ masses. In practice, because of an
easier experimental measurement, we have only considered the 
effects on the $Z_L$ rate (and not shown the individual $W^{\pm}_L$ ones).\\ 

We have made illustrations showing the mentioned effects
and their specificities (scale dependent masses, final state 
interactions) on the $p_Z$
distribution of the $Z_L$ rate in the $e^+e^-\to ZWW $ process.
For experimental possibilities see \cite{ee}.\\

Other production processes may also be interesting for our aim.
One set is
$ZWW$ production from different initial states like $\gamma-\gamma$, see \cite{gammagamma},
or gluon-gluon in hadronic collisions; for LHC possibilities see 
\cite{lhcContino,lhcRichard}.\\
Another possibility is $HWW$ or $HZZ$ production which will
also directly depend on $Z,W$ masses but the identification of the $H$
may be more delicate than for a $Z$.\\

\newpage

\begin{figure}[p]
\vspace{-4cm}
\[
\hspace{-2cm}\epsfig{file=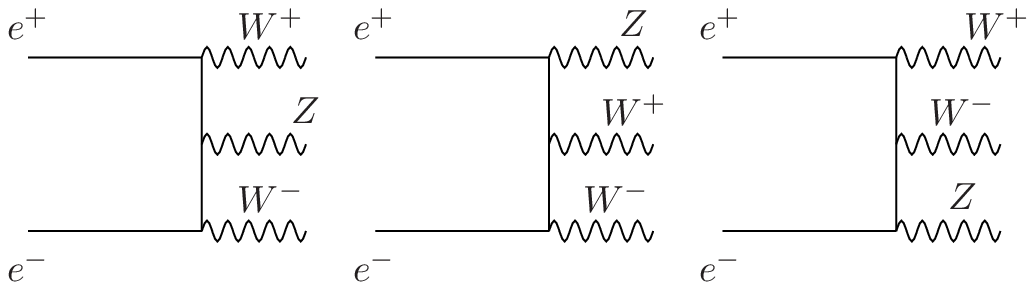 , height=2.5cm}
\]\\

\vspace{-0.5cm}
\[
\hspace{-2cm}\epsfig{file=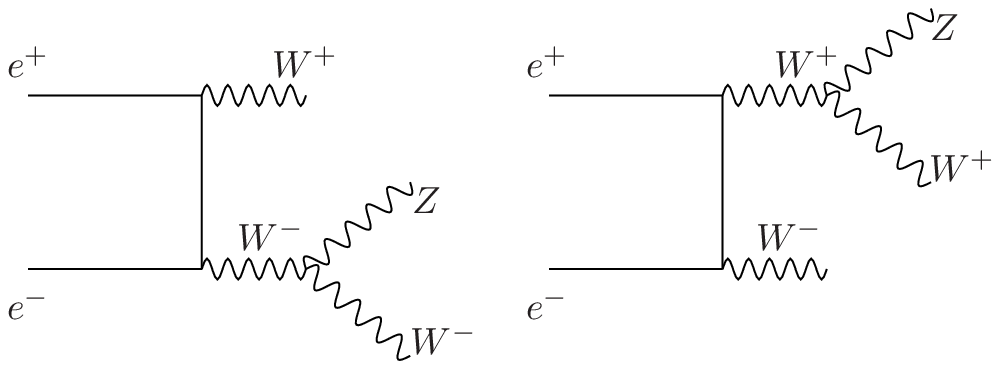 , height=2.5cm}
\]\\
\vspace{-0.5cm}
\[
\hspace{-2cm}\epsfig{file=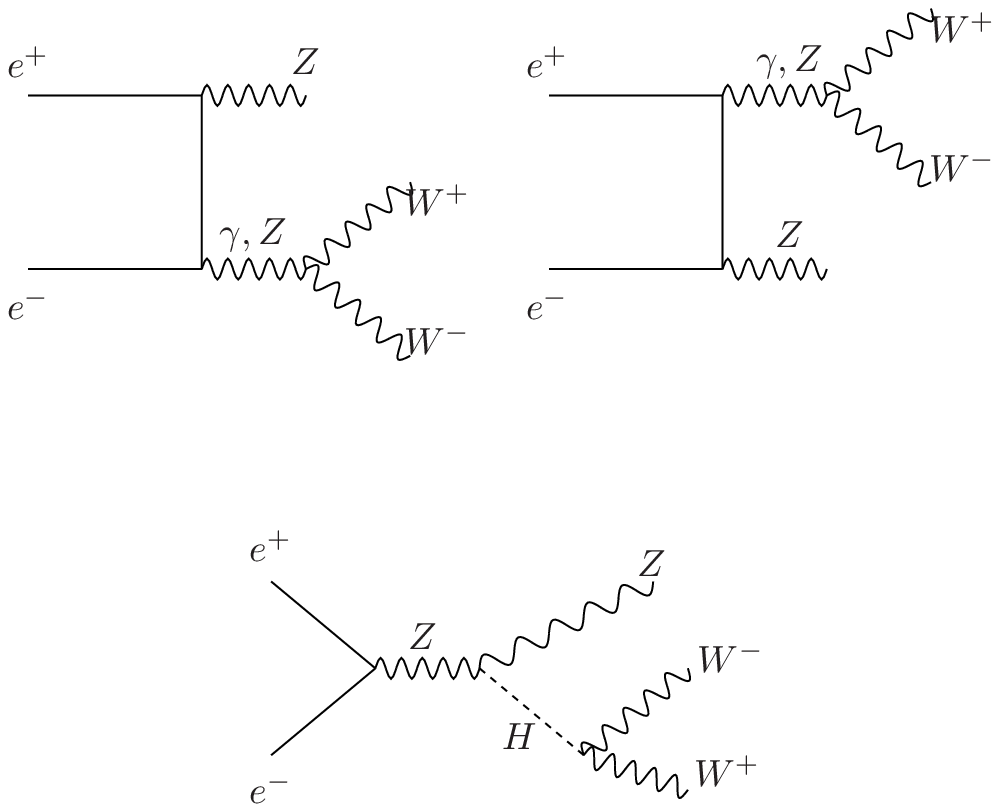 , height=5.cm}
\]\\
\vspace{-1cm}
\[
\hspace{-2cm}\epsfig{file=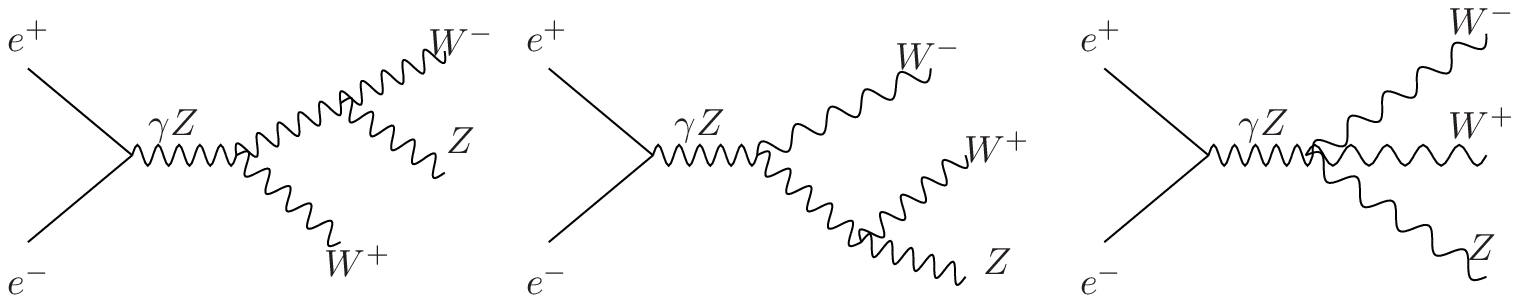 , height=2.5cm}
\]\\

\caption[1] {Diagrams for $e^+e^- \to ZWW$ .}
\end{figure}
\clearpage

\begin{figure}[p]
\vspace{-0cm}
\[
\hspace{-2cm}\epsfig{file=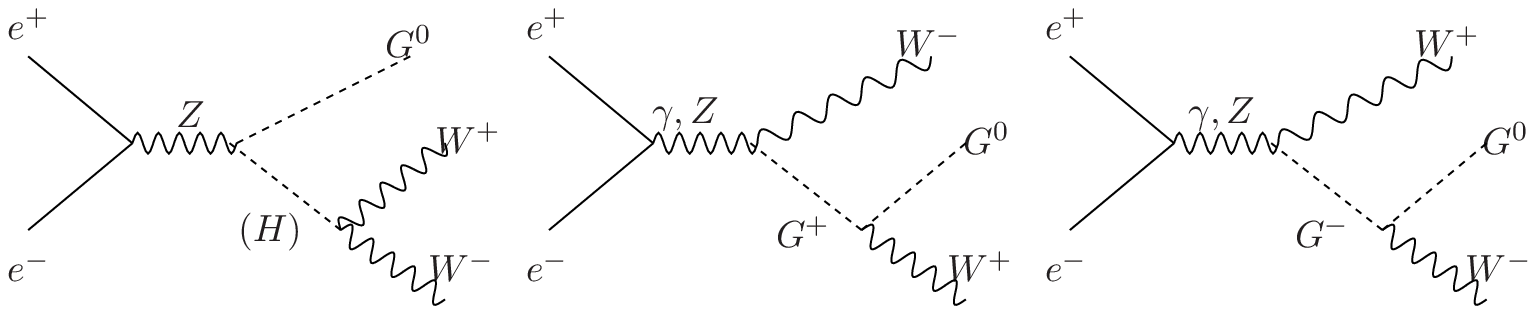 , height=3.2cm}
\]\\
\vspace{-1cm}
\hspace{0cm}(a)\hspace{6cm}(b)\hspace{5cm}(c)

\vspace{2cm}
\caption[1] {Diagrams for $e^+e^- \to G^0WW$ .}
\end{figure}

\clearpage

\begin{figure}[p]
\vspace{-0cm}
\[
\hspace{-2cm}\epsfig{file=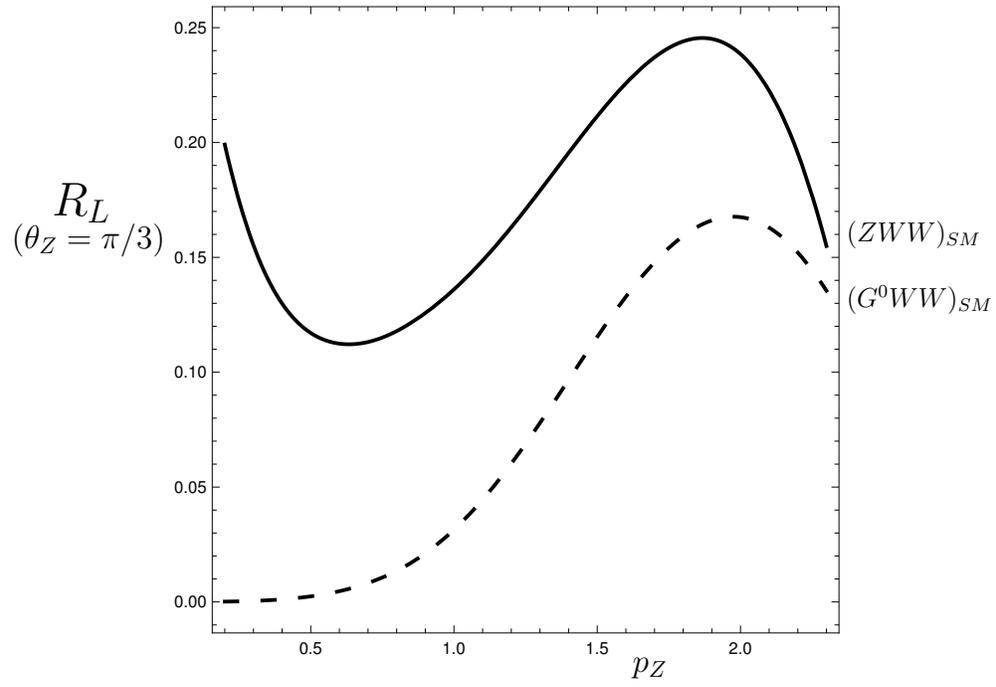 , height=9.cm}
\]\\
\vspace{0.5cm}
\vspace{-0cm}
\[
\hspace{-2cm}\epsfig{file=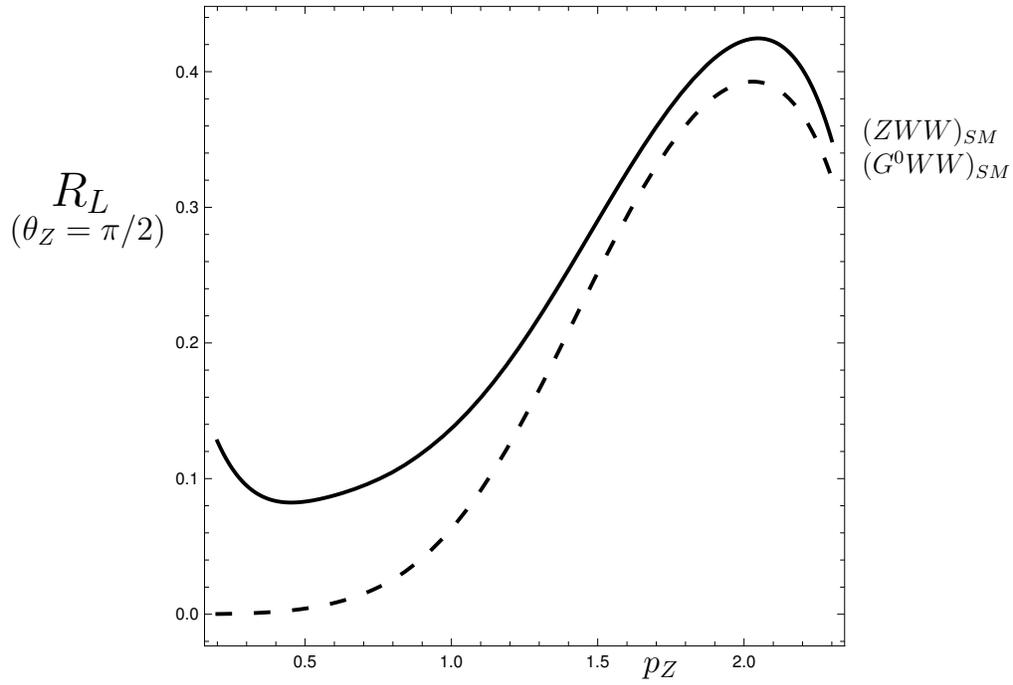 , height=9.cm}
\]\\
\vspace{-1cm}
\caption[1] {$e^+e^- \to Z_LWW $ ratio in SM case compared to the
Goldstone $G^0WW $ case.}
\end{figure}

\clearpage
\begin{figure}[p]
\vspace{-0cm}
\[
\hspace{-2cm}\epsfig{file=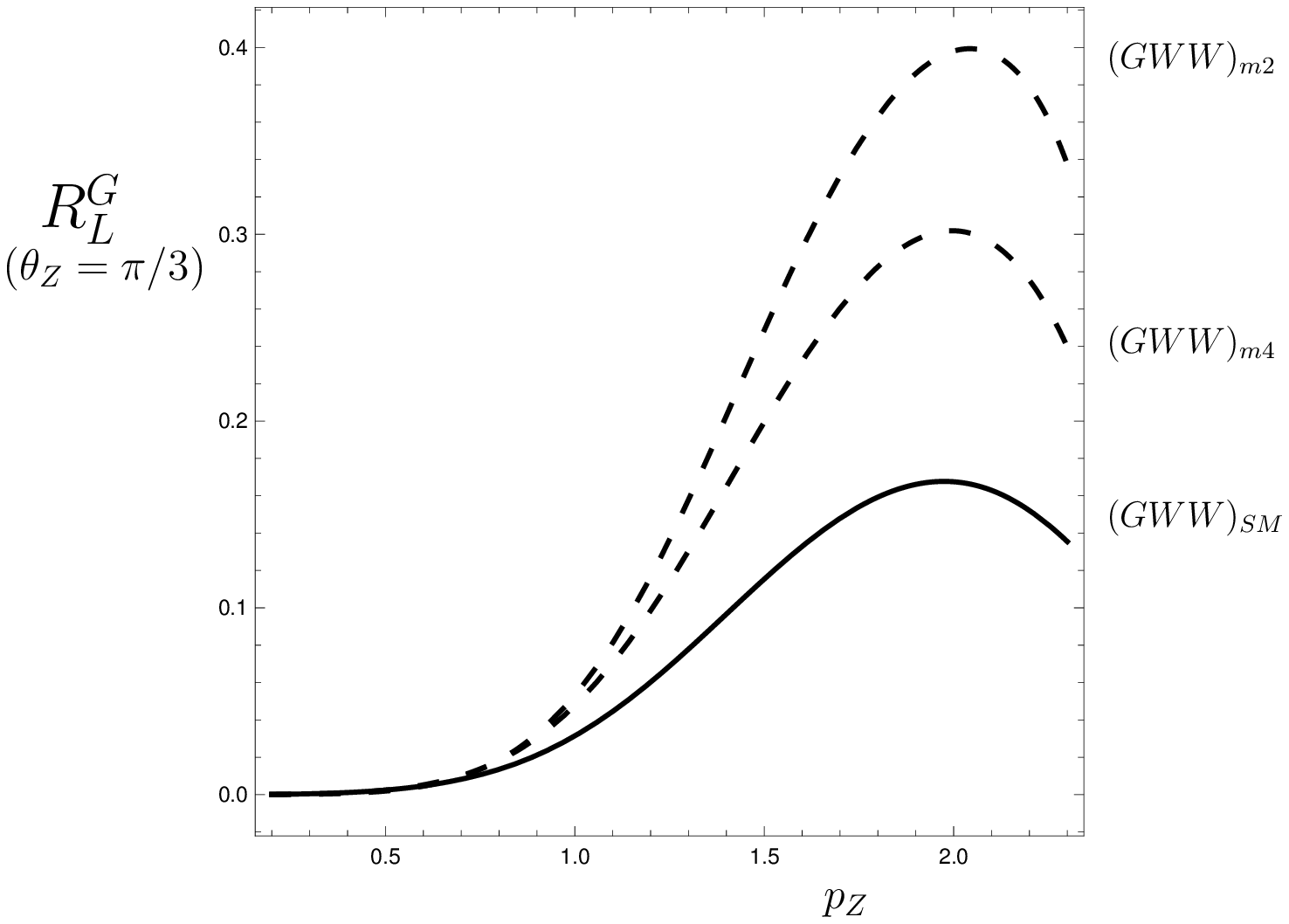 , height=9.cm}
\]\\
\vspace{0.5cm}
\vspace{-0cm}
\[
\hspace{-2cm}\epsfig{file=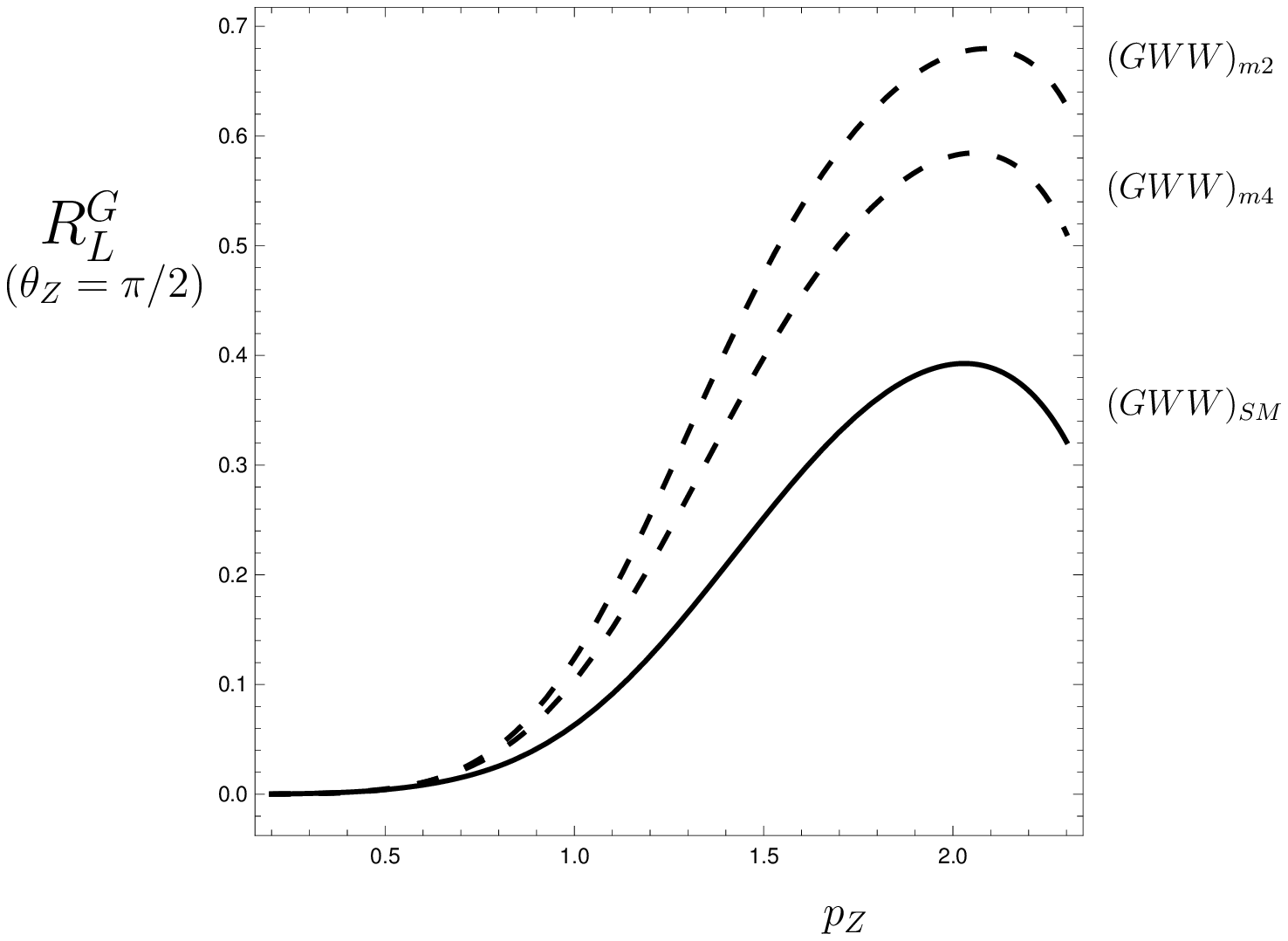 , height=9.cm}
\]\\
\vspace{-1cm}
\caption[1] {$e^+e^- \to G^0WW  $ ratio for 2 cases of scale dependent mass compared to the SM case.}
\end{figure}

\clearpage
\begin{figure}[p]
\vspace{-0cm}
\[
\hspace{-2cm}\epsfig{file=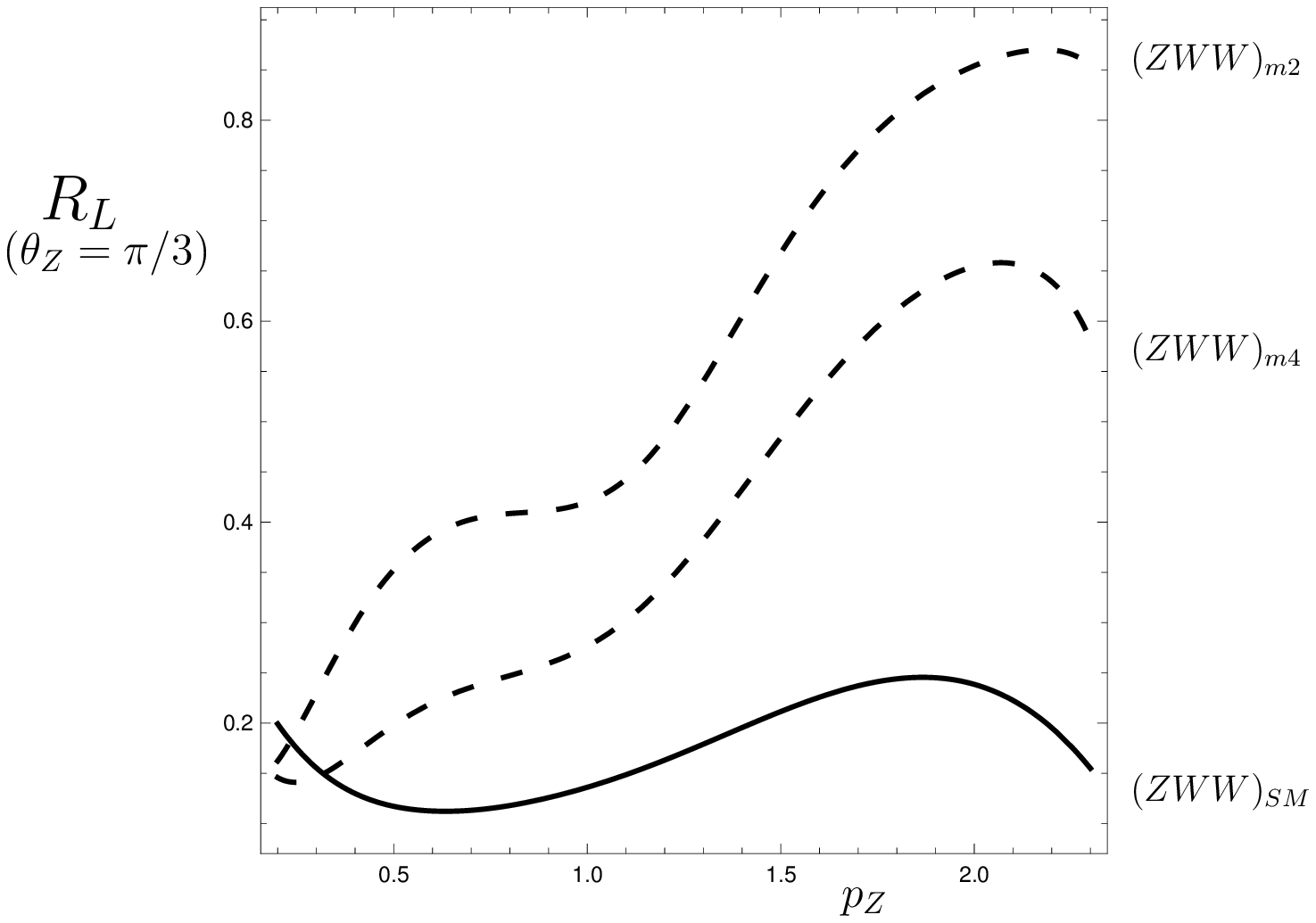 , height=9.cm}
\]\\
\vspace{0.5cm}
\vspace{-0cm}
\[
\hspace{-2cm}\epsfig{file=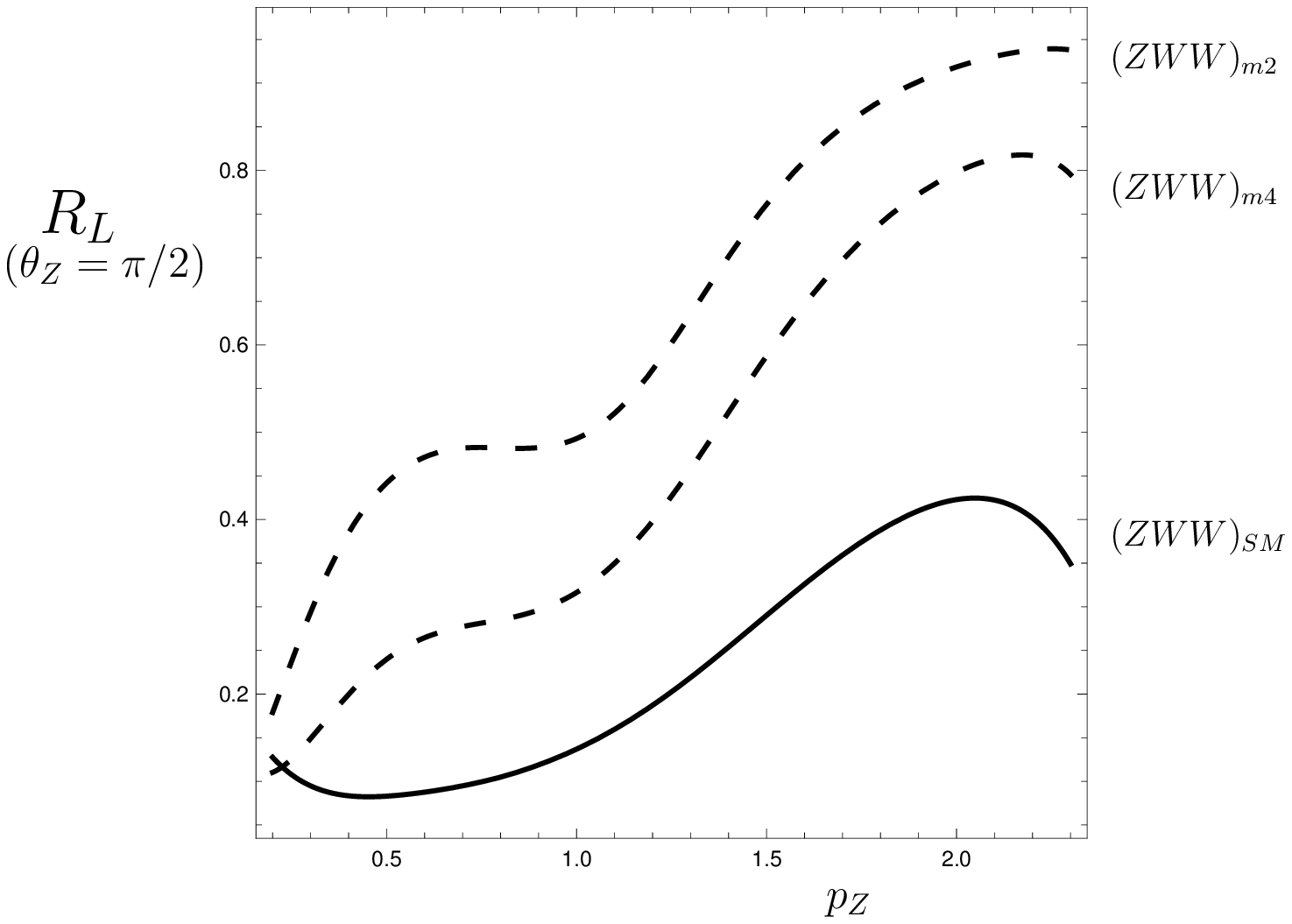 , height=9.cm}
\]\\
\vspace{-1cm}
\caption[1] {$e^+e^- \to Z_LWW  $ ratio for 2 cases of scale dependent mass compared to the SM case.}
\end{figure}

\clearpage
\begin{figure}[p]
\vspace{-0cm}
\[
\hspace{-2cm}\epsfig{file=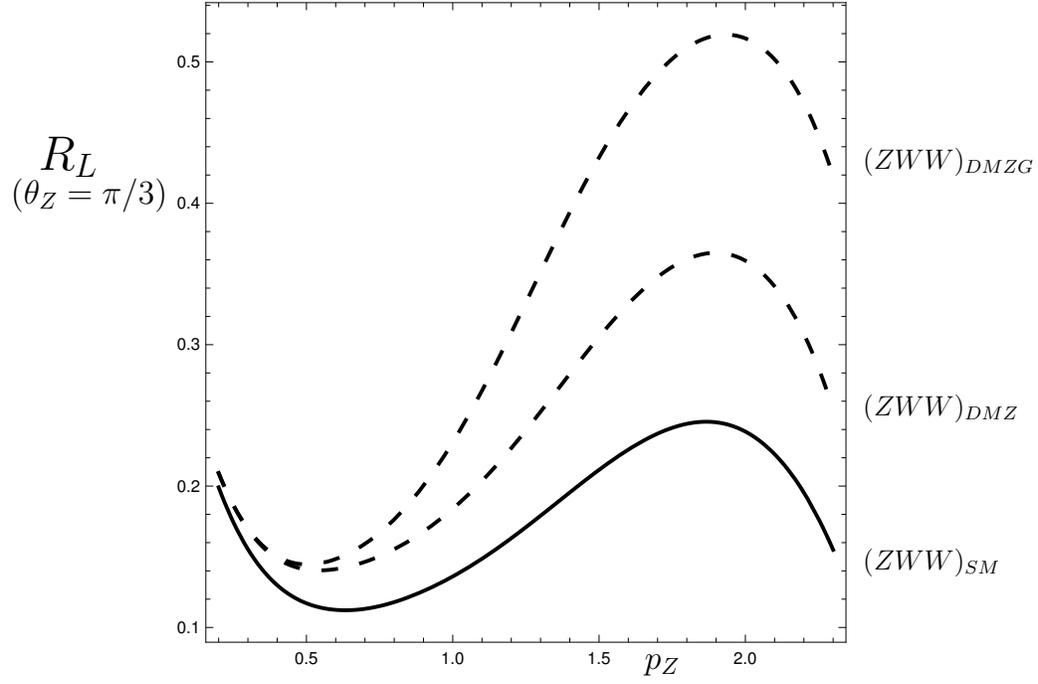 , height=9.cm}
\]\\
\vspace{0.5cm}
\vspace{-0cm}
\[
\hspace{-2cm}\epsfig{file=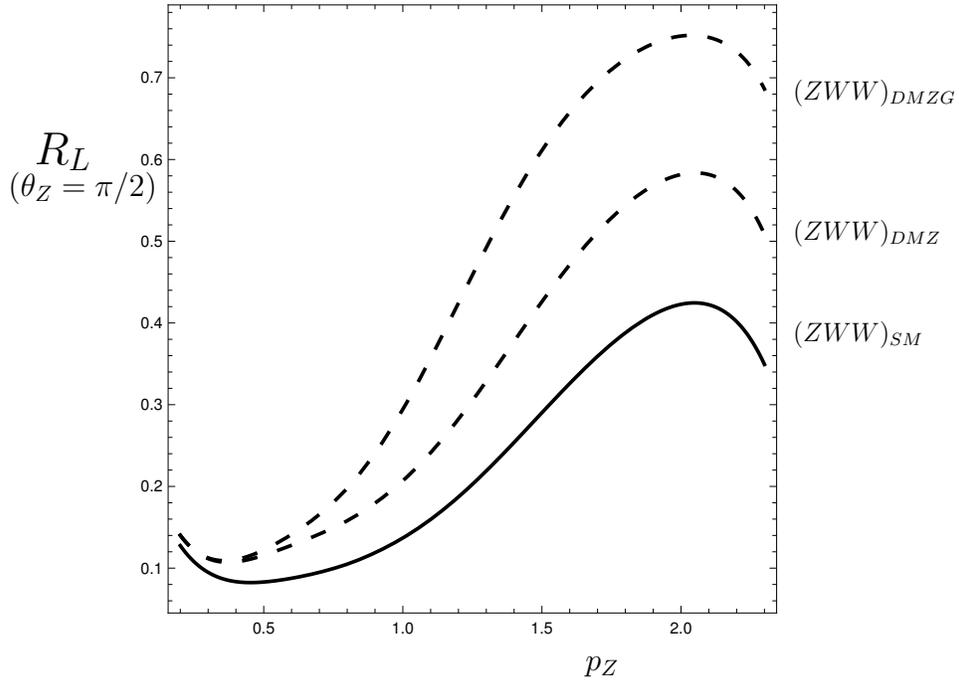 , height=9.cm}
\]\\
\vspace{-1cm}
\caption[1] {$e^+e^- \to Z_LWW  $ ratio for 2 cases of final state interaction compared to the SM case.}
\end{figure}

\end{document}